\documentclass[aps, prd, twocolumn, preprintnumbers, letterpaper, nofootinbib, superscriptaddress, balancelastpage, 10pt]{revtex4-2}
\usepackage{cancel}
\usepackage[utf8]{inputenc}
\usepackage{graphicx}
\usepackage{hyperref}
\usepackage{xcolor}
\usepackage{amsmath, amsfonts, amssymb}
\usepackage{bm, bbm}
\usepackage{lipsum}
\usepackage{enumitem}
\usepackage[caption=false]{subfig}
\usepackage{slashed}
\usepackage{tikz-feynman}
\usepackage[normalem]{ulem}
\newcommand{\beq}{\begin{equation}\begin{aligned}{}}
\newcommand{\eeq}{\end{aligned}\end{equation}}
\newcommand{\beqa}[1]{\begin{equation}\begin{aligned}{#1}}
\newcommand{\eeqa}{\end{aligned}\end{equation}}

\newcommand{\bea}{\begin{eqnarray}{}}
\newcommand{\eea}{\end{eqnarray}}

\begin{document}

\title{Photo-production of axions in Supernovae}

\author{Sabyasachi Chakraborty}
\email{sabyac@iitk.ac.in}
\affiliation{Department of Physics, Indian Institute of Technology, Kanpur-208016, India}

\author{Aritra Gupta}
\email{aritra.gupta@ific.uv.es}
\affiliation{Instituto de F\'isica Corpuscular (IFIC), CSIC,
Parc Cient\'ific, C/Catedr\'atico Jos\'e Beltr\'an, 2, E-46980 Paterna, Spain}

\author{Miguel Vanvlasselaer}
\email{miguel.vanvlasselaer@vub.be}
\affiliation{Theoretische Natuurkunde and IIHE/ELEM, Vrije Universiteit Brussel,
\& The International Solvay Institutes, Pleinlaan 2, B-1050 Brussels, Belgium}

\begin{abstract}
Compact stellar objects like supernovae and neutron stars are believed to cool by emitting axions predominantly via axion bremsstrahlung ($NN \to NNa$), pion conversion ($\pi^- p^+ \to N a$) and photo-production ($\gamma N \to N a$). In this paper, we study in detail the photo-production channel, from the unavoidable anomaly induced Wess-Zumino-Witten term $\propto \epsilon^{\mu \nu \alpha \beta}\, F_{\mu \nu}\, \partial_\alpha a \, \omega_\beta$ in conjunction with the low energy pion photo-production data.  We found that for heavier axions, i.e., $m_a\sim\mathcal{O}(100)$ MeV, photo-production processes can be dominant compared to the usual axion emission processes. In addition, the spectrum of axions emitted in the process is significantly harder than those originating from bremsstrahlung.
\end{abstract}

\maketitle

\section{Introduction} Searches for new physics much lighter than the electroweak scale have gained a lot of attention in the recent past. Among all the possible candidates, axions or axion-like particles are perhaps the most well-motivated. Axion emerges as the pseudo-Nambu-Goldstone boson of a spontaneously broken $U(1)$ symmetry~\cite{Peccei:1977hh,Peccei:1977ur,Weinberg:1977ma,Wilczek:1977pj} and offers a compelling solution to the long-standing strong-$CP$ problem~\cite{tHooft:1976rip} of the Standard Model (SM). Moreover, axions can also be a natural dark matter candidate~\cite{Preskill:1982cy,Dine:1982ah,Abbott:1982af}, address the hierarchy problem~\cite{Graham:2015cka, Hook:2016mqo, Trifinopoulos:2022tfx} and play a crucial role in resolving the matter-antimatter asymmetry~\cite{Co:2019wyp,Chakraborty:2021fkp} of the Universe. Naturally, axion furnishes complementary means to probe physics beyond the SM at multiple frontiers. 

While the axion can have a myriad of couplings with the SM particles, the most minimal effective Lagrangian addressing the strong $CP$ problem, up to terms required for renormalization is given as:
\begin{eqnarray}
    \mathcal L = \mathcal L_{\text{SM}} + \frac{\alpha_s}{8\pi} \frac{a}{f_a} G\widetilde{G} + \frac{1}{2} \left(\partial_\mu a\right)^2 - \frac{1}{2} m_{a_0}^2 a^2\;,
    \label{eq:axion_Lag}
\end{eqnarray}
where $G$ is the field strength tensor of the gluon and $a$ is the axion with decay constant $f_a$. For lighter axions, $m_a\lesssim\mathcal{O}(\text{GeV})$, it is convenient to rotate away the $aG\widetilde{G}$ coupling by a chiral transformation of the SM quark fields. The ensuing Lagrangian generates axion-quark kinetic and mass mixing terms and can be matched with an effective theory such as Chiral perturbation theory ($\chi$PT) containing mesons. This effective description provides a powerful framework to study non-perturbative effects such as axion-pion mixing and has been used to constrain large parts of axion parameter space from intensity frontier experiments such as beam dump~\cite{Bjorken:1988as,Blumlein:1990ay,CHARM:1985anb}, rare decays of pions~\cite{PIENU:2019usb,Pocanic:2003pf,Altmannshofer:2019yji} and kaons~\cite{Georgi:1986df,Bardeen:1986yb,Alves:2017avw,Gori:2020xvq,E949:2005qiy,NA62:2014ybm,KOTO:2018dsc,Bauer:2021wjo}, etc. However, for heavy QCD axions, $m_a\gtrsim\mathcal{O}(1)$ GeV, the power counting of $\chi$PT breaks down and one generally relies on pertubative computations to probe axions~\cite{Aloni:2018vki,Chakraborty:2021wda,Bertholet:2021hjl,Bauer:2021mvw}, mostly from rare $B$-decays.

On the other hand, the cosmic frontier provides complementary means to probe axions as it is sensitive to lighter masses, e.g., $m_a\lesssim\mathcal O(100)$ MeV. For example, nuclear reactions or thermal processes inside the stellar interior such as White Dwarf (WD), Neutron Star (NS), Supernova (SN), etc., are potentially powerful sources of axions. The emission of axions from such compact stellar objects might result in a more efficient transport of energy compared to the SM neutrinos, leading to observational changes. This lead the authors of Ref.~\cite{1990PhR...198....1R,Raffelt:1996wa} to propose the following bound on the emissivity of axions 
\bea
\label{Eq:bound_WZW}
    \frac{Q_a}{\rho} \lesssim 10^{19}\; \text{erg}\; s^{-1} g^{-1} \;,
\eea
evaluated at a temperature $T=30-50$ MeV and around the nuclear saturation density. Traditionally, axion bremsstrahlung, i.e.,  $NN \to NNa$~\cite{PhysRevLett.53.1198, Iwamoto:1992jp}, where $N= (p,n)$ is a nucleon, was considered to be the most dominant channel of axion production in the core of the SN. Although previous studies assumed One Pion Exchange (OPE) approximation, large suppression was found while going beyond OPE \cite{Hanhart:2000ae}. Nevertheless, using Eq.\eqref{Eq:bound_WZW}, earlier bounds were drawn on the axion mass and effective axion-nucleon coupling~\cite{Fischer:2016cyd,PhysRevD.42.3297,PhysRevLett.65.960, PhysRevD.56.2419, MAYLE1989515, PhysRevD.40.299, Ho:2022oaw}. More recently, it was argued that pion conversion~\cite{PhysRevD.45.1066, PhysRevD.52.1780} could potentially enhance the emissivity by a factor of few~\cite{PhysRevLett.126.071102, Choi:2021ign}, so is the existence of quark matter in the core of the SN~\cite{Cavan-Piton:2024ayu}. We summarize those contributions and the relevant operators in Table \ref{tab_cross_sec}.

\begin{table}[h!]
\begin{center}
\begin{tabular}{lll}
\hline
\multicolumn{1}{|l|}{Process} & \multicolumn{1}{l|}{Coupling} & \multicolumn{1}{l|}{Refs}  \\ \hline
\multicolumn{1}{|l|}{$NN \to NNa$} & \multicolumn{1}{l|}{$\left(C_{aN}/2f_a\right)\; \bar N \gamma_5 \gamma_\mu N\; \partial^\mu a$} & \multicolumn{1}{l|}{\cite{PhysRevLett.53.1198, Iwamoto:1992jp}} 
\\ \hline
\multicolumn{1}{|l|}{$\pi^- p \to Na$} & \multicolumn{1}{l|}{$\left(C_{aN}/2f_a\right)\; \bar N \gamma_5 \gamma_\mu N\; \partial^\mu a$} & \multicolumn{1}{l|}{\cite{PhysRevD.56.2419,PhysRevLett.126.071102}} 
\\ \hline
\multicolumn{1}{|l|}{$N\gamma \to Na$} & \multicolumn{1}{l|}{$i\;\left(C_{aN\gamma}/2\right) a\; \bar N \gamma_5 \sigma_{\mu \nu} N\; F^{\mu \nu}$} & \multicolumn{1}{l|}{\cite{Graham:2013gfa,Lucente:2022vuo}} 
\\ \hline
\multicolumn{1}{|l|}{$N\gamma \to Na$} & \multicolumn{1}{l|} {$\left(\kappa/f_a\right)\epsilon^{\mu \nu \rho \sigma}\; F_{\mu \nu}\;  \partial_\rho a\; \omega_\sigma$} & \multicolumn{1}{l|}{This work}
\\ \hline
\end{tabular}
\end{center}
\caption{Interactions relevant for axion emission from a supernova. The coefficient $\kappa$ is defined in Eq.~\eqref{eq:axion_WZW}.
\label{tab_cross_sec}} 
\end{table}

In this minimal scenario, it is natural to ask whether there exist any other unavoidable interactions of axions or not. For example, in the SM, there exists Wess-Zumino-Witten (WZW) interactions~\cite{Wess:1971yu,Witten:1983tw,Kaymakcalan:1983qq,Chou:1983qy,Kawai:1984mx,Pak:1984bn,Harvey:2007ca,Harvey:2007rd,Hill:2007zv,Chakraborty:2023wgl} which can account for certain processes including anomalies that can not be generated in the usual $\chi$PT. This is because low-energy effective theories such as $\chi$PT  can often possess more symmetries such as spurious parity than its UV counterpart QCD. Interestingly, in the presence of background gauge fields (such as $\omega$, $\rho$ mesons, etc.), the physics of WZW interactions are quite rich~\cite{Harvey:2007rd}. Recently it was shown that even in SM, such interactions open up new channels for the cooling of young neutron stars~\cite{Harvey:2007ca, Chakraborty_2023, Bai:2023bbg}. In this paper, we study the implications of photo-production $\gamma n \to n a$ and WZW terms in the presence of axions leading to novel interactions. We also compare our process with the usual photo-production channels.

 The paper is organized as follows. In section~\ref{section:photo-production} we discuss in detail the different processes pertaining to axion photo-production. After briefly reviewing the widely used dipole operator, we point out the novel Wess-Zumino-Witten interaction which results in the photon initiated axion production in neutron rich stellar objects. We compute the emissivity for both degenerate and non-degenerate environments and found that this contribution is sub-leading compared to the usual processes. Finally, we also point out a complementary way of estimating the total emissivity from an entirely data-driven approach.
\begin{figure}
   \centering    
\includegraphics[width=\linewidth]{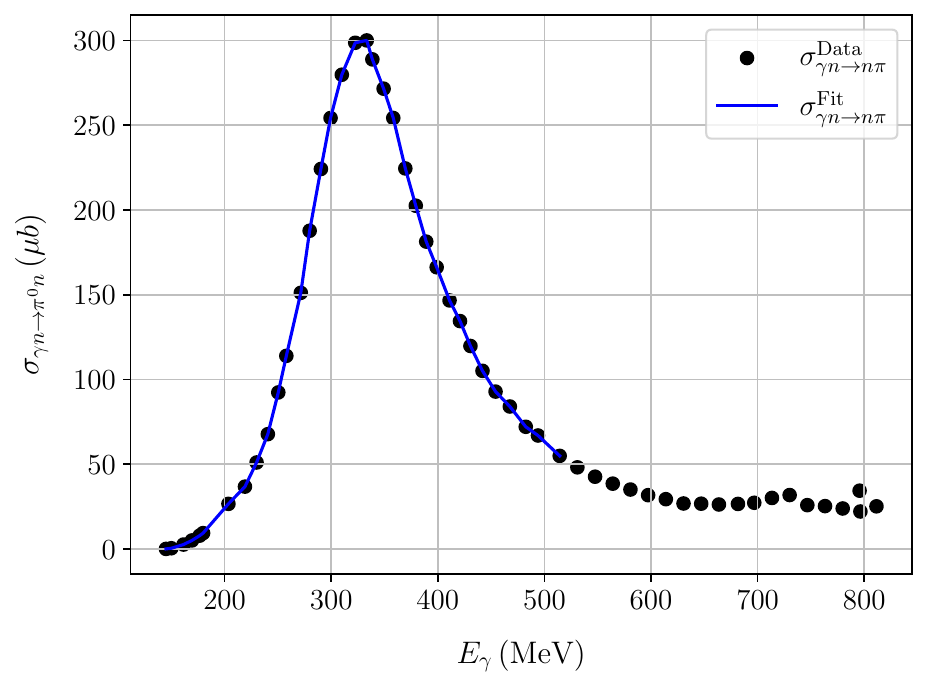}
   \caption{Cross-section of the $\gamma p \to p a$ against the data from the  PIONS@MAX-lab Collaboration~\cite{Strandberg:2018djk} presented in \cite{Briscoe:2020qat,A2:2019yud,Drechsel:2007if}. In blue we present a fit on the data that we used for computing the emissivity. }
    \label{fig:FIT}
\end{figure}
Using low energy data of photon initiated pion production (see fig.~\ref{fig:FIT}), we estimate the overall axion photo-production rate. Interestingly, this model independent study provides a comparable limit on the axion parameter space, for $m_a\sim \mathcal O(100)$ MeV. We also point out the hardness of the emitted axion spectrum and comment on its prospect of detection. Finally, in section~\ref{sec:conclusion}, we conclude.

\section{Photo-Production of Axions}
\label{section:photo-production}
In this section, we provide a comparative study of the usual dipole induced axion photo-production channel with the novel WZW interaction and data driven approach.
\subsection{Dipole induced photo-production:} Photo-production of axions can occur via the direct coupling of axions to nucleons and photons. This is usually induced through the unavoidable electric dipole portal, such as $i\;\left(C_{aN\gamma}/2\right) a\; \bar N \gamma_5 \sigma_{\mu \nu} N\; F^{\mu \nu}$. Implications of such an interaction have been studied in depth in\cite{Lucente:2022vuo}, however, the bounds obtained on the axion decay constant are roughly around $f_a/\text{GeV}\gtrsim 6 \times 10^5 $. This is indeed subleading compared to the prototypical axion cooling channels and even other photo-production processes stemming from the WZW and data driven contribution. As a consequence, we will neglect the possible cross terms between the other channels and the electric dipole that could arise. Nevertheless, dipole induced interactions can have interesting signatures in future neutrino experiments.

\subsection{Wess-Zumino-Witten induced photo-production:} To generate WZW interactions, one considers the 5-dimensional action~\cite{Wess:1971yu,Witten:1983tw} which is invariant under chiral symmetry, where the boundary is identified with our 4-dimensional spacetime, e.g.,
\begin{eqnarray}
\label{eq:WZW_U2}
&& S_{\rm WZW} (A_\mu, U) = N_c \int_D\; d^5 y\; \omega\;, \nonumber \\
&&     \omega = -\frac{i}{240\pi^2}\;  \epsilon^{\mu\nu\rho\sigma\tau}\;\text{Tr}\left(\mathcal{U}_\mu\;\mathcal{U}_\nu\;\mathcal{U}_\rho\; \mathcal{U}_\sigma\;\mathcal{U}_\tau\right)\;.
    \label{eq:WZW_U1}
\end{eqnarray}
Here $\mathcal{U}=U^{\dagger}\partial_\mu U$ with $U=\text{Exp}\left(2i\pi^a T^a/f_\pi\right)$. $\pi_a$ are the pion fields with decay constant $f_\pi$ and $A_\mu$'s are the gauge fields. The coefficient $N_c=3$ is fixed by matching with QCD. Any arbitrary subgroup of the chiral symmetry can be gauged but only in the 4-dimension using the trial and error method. The full result has been nicely tabulated in~\cite{Witten:1983tw} (also see~\cite{Wess:1971yu,Kaymakcalan:1983qq,Chou:1983qy,Kawai:1984mx,Pak:1984bn}). Meson fields such as $\omega$ can be introduced as a background vector field as prescribed in Ref.~\cite{Harvey:2007ca} by replacing $\tilde{A_\mu}=A_\mu+B_\mu$ in the effective action. If the fundamental gauge fields are vector-like, i.e., $A_L=A_R$, one needs to add appropriate counter-terms to maintain gauge invariance and conservation of vector current. Finally, tracking interactions with the pion fields (see Appendix \ref{App:WZW} for details), we find

    \begin{align}
    \hspace*{-0.24cm}
         \mathcal{L}_{WZW}^{\pi_0} & \supset \; \frac{N_c\; \epsilon^{\mu\nu\rho\sigma}}{24\pi^2} F_{\mu\nu}\left[\frac{e^2}{4}\frac{\pi_0}{f_\pi}\;F_{\rho\sigma}+eg_\omega\frac{\partial_\rho \pi_0}{f_\pi}\; \omega_\sigma \right]\;.
         \label{eq:pion_WZW}
    \end{align}

Starting from Eq.~\eqref{eq:axion_Lag}, axions can be incorporated in this framework by a suitable chiral transformation on the SM quark fields. This generates a mass-mixing between the axion and pion fields (shown by a cross in Fig.~\eqref{fig:Cooling_diag}). The mixing angle can be approximated as\cite{DiLuzio:2020wdo,Notari:2022ffe,GrillidiCortona:2015jxo} 
\begin{equation}
\theta_{\pi^0-a} \simeq \frac{f_\pi}{2 f_a} \bigg(\frac{m_d-m_u}{m_u+m_d} \bigg) \equiv \frac{f_\pi}{f_a} C_A \,.
\label{eq:mixing}
\end{equation}

We then trade the mixing angle between axion and pion fields, which generates axion-photon coupling and the desired interactions of the form
\begin{equation}
\mathcal L_{\rm WZW}^a \supset \frac{\kappa}{f_a}\; \epsilon^{\mu\nu\rho\sigma}\; F_{\mu\nu}\;\partial_\rho a \; \omega_\sigma\; , \; \kappa = \frac{C_A N_c}{24\pi^2}\; e g_\omega\;.
    \label{eq:axion_WZW}
\end{equation}
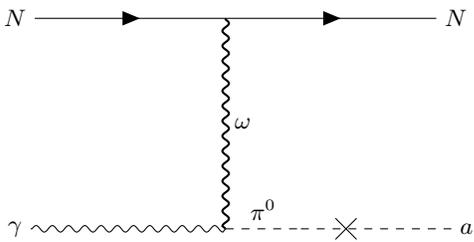
\begin{figure}[t!]
\center
\begin{tikzpicture} 
\begin{feynman}
\vertex (a1) {\(N\)}; 
\vertex[right=2.8cm of a1] (a2); 
\vertex[right=2.8cm of a2] (a3) {\(N\)}; 
\vertex[below=2.8cm of a1] (b1) {\(\gamma\)}; 
\vertex[right=2.8cm of b1] (b2) ;
\vertex[right=1 cm of b2] (b3); 
\vertex[right=2.0cm of b3] (g){\(a\)}; 
\diagram*{
(a1) -- [fermion] (a2),
(a2) --[fermion] (a3),
(b1)--[boson] (b2),
(b2)--[scalar, edge label= {\(\pi^0\)}] (b3),
(a2)--[boson, thick, edge label=\(\omega\)](b2),
(b3)--[scalar, insertion={[size=4pt]0.3}](g)
};
\end{feynman} 
\end{tikzpicture}
 \caption{Cooling of supernovae and neutron stars via $N \gamma \to Na$ induced by axion-WZW interactions. The cross indicates pion-axion mixing.}
\label{fig:Cooling_diag}
\end{figure}
On the other hand, nucleons interact with the vector meson fields $\omega$ via
\begin{equation}
\mathcal{L}_0 = \bar{N}\left(i\cancel{\partial}-g_\omega\;\cancel{\omega}-m_N\right) N\;,
\label{eq:1st_Lag}
\end{equation}
where $g_\omega$ is estimated using nuclear physics and its value depends on the fitting model at hand. It is however estimated to be roughly $g_\omega \approx 10-12$~\cite{BRACCINI1970173}. 
More generally, from the total cross-section for the pion production, we can obtain the rate for the axion production 
\bea 
\sigma_{\gamma n \to n a} = \theta_{\pi^0-a}^2\sigma_{\gamma n \to n \pi_0} \, ,
\eea 
in the regime $E \gg m_a, m_\pi$. 
 
\subsubsection{Computing the emissivity} As depicted in Fig.~\ref{fig:Cooling_diag}, Eq.~\eqref{eq:axion_WZW} and Eq.~\eqref{eq:1st_Lag} provide an efficient production channel of axions in a nucleon rich environment. The coupling $f_a$, dictates whether the axions can come out of the stellar object or get trapped inside it. In the former case, axions can carry a part of the internal energy of the core with them, resulting in the cooling of the star. To estimate, we compute the emissivity $Q$, defined as the emitted energy per unit volume and per unit time. We take into account both neutron degenerate (D) as well as non-degenerate (ND) scenarios, mimicking the situation inside an NS and SN respectively. The emissivity for axion photo-production $\gamma N\to N a$ is given by
\begin{align}
\label{eq:emissivity_exp}
& Q_{N\gamma \to N a} \equiv \int \frac{d^3 \vec{p}_\gamma f_\gamma\left(p_\gamma\right)}{(2\pi)^3 2E_\gamma} \int \frac{d^3  \vec{p}_{N_1}}{(2\pi)^3 2E_{N_1}}\frac{d^3  \vec{p}_{N_2}}{(2\pi)^3 2E_{N_2}} \nonumber \\
&g_\gamma \,g_N \int \frac{d^3 \vec{p}_a}{(2\pi)^3 2E_a} E_a \langle |\mathcal{M}|^2\, \rangle   f_N(E_{N_1})\,(1-f_N(E_{N_2}))\nonumber \\
&\quad\times \, (2\pi)^4 \delta (E_{N_1}- E_{N_2}-Q_0)\,\delta^3(\vec{p}_{N_1}-\vec{p}_{N_2}- \vec{q})\;,  
\\ \nonumber
&\approx   \int \frac{g_\gamma d^3 \vec{p_\gamma}}{(2\pi)^3 }  f_\gamma (p_\gamma) \int \frac{ g_N d^3 \vec{p}_{N}}{(2\pi)^3 }  f_N \sigma_{\gamma N \to N a} (E_\gamma) E_a\;, 
\end{align}
where $\sigma(E_\gamma)$ is the integrated cross section. 
where $g_{\gamma(N)} =2$ are the dof. of the photon (neutron), $f_N$ is the neutron distribution function, $E_i$ and $\vec{p}_i$ are the energies and three-momenta of the $i^{\rm th}$ species and $Q_0 \equiv E_a-E_\gamma,\; \vec{q} \equiv \vec{p}_a-\vec{p}_\gamma$. The kinematics of the process is similar to a fixed target experiment where the initial neutron is assumed to be at rest. Due to the momentum transfer, the final neutron receives a kick and we consider them to be nearly non-relativistic thereby expanding their energies as $E_{N_i} \approx m_N + \mathcal{O}(\vec{p}^2_{N_i}/2m_N)$. The nucleon density is given by 
\begin{equation}
\label{eq:density_neu}
 n_B \equiv \int  \frac{d^3 p_{N}}{(2\pi)^3  }  g_N  f_N ,  \qquad n_B = 4.5 \times  10^{6} \rho_{15} \text{ MeV}^3  \,. \nonumber \\
\end{equation}
Thus the emissivity from photo-production finally simplifies to
\begin{align}
\label{eq:emissivity_exp}
 Q_{N\gamma \to N a} \approx   \int \frac{ d E_\gamma }{\pi^2 } E_\gamma^3 f_\gamma (p_\gamma) n_B \sigma_{\gamma N \to N a} (E_\gamma) \;, 
\end{align}

We now turn to the computation of the emissivity using Eq.\eqref{eq:emissivity_exp}. 

\subsubsection{Emissivity from WZW term alone}
We first compute the emissivity from the WZW term only. The spin averaged cross section reads  
\begin{align} 
 \sigma^{\rm WZW}(E_\gamma) \approx \frac{1}{3 \pi } \frac{\kappa^2 g_\omega^2}{f_a^2 m_\omega^4} \sqrt{1 - \frac{m_a^2}{E_\gamma^2}}   E_\gamma^2 (E_\gamma^2 - m_a^2) \;. 
\end{align}
Here, $\theta$ is the angle between the axion and photon three momenta. The kinematics of the process dictates $E_a\sim E_\gamma$ when $m_\gamma \ll T$. This is a reasonable approximation as in the core of SN, the photons acquire a mass of the order of $m_\gamma \sim \mathcal O(10)\; \text{MeV} \times (Y_e\rho_{14})^{1/3}$, where $Y_e$ is the electron fraction. Since $T/m_\gamma \sim (3-4)$, we therefore do not expect the mass of the photon to play a significant role. 

\subsubsection{Non-degenerate scenario} For ND neutrons, the absence of Pauli blocking helps us to further simplify Eq.~\eqref{eq:emissivity_exp} as $1-f_N(p_N)\approx 1$. Using the thermal distribution for photons~\cite{Payez:2014xsa} and integrating the axion energy from $(m_a, \infty)$, we find 
\begin{align}
    Q_{N \gamma \to N a}^{\rm WZW, ND} \approx  & \; g_\gamma g_\omega^2\;  \frac{\kappa^2 n_b}{2\pi^3}\; \frac{T^2}{f_a^2}\; \frac{m_a^4}{m_\omega^4}\; \left[4m_a T\; K_3\left(\frac{m_a}{T}\right) \right.\nonumber\\
    &\left. +(m_a^2+35T^2)\;K_4\left(\frac{m_a}{T}\right)\right]\;,
    \label{eq:ND}
\end{align}
where $K_{3,4}$ are the Bessel function of type three and four respectively. In the limit of $m_a\to 0$, the expression in Eq.\eqref{eq:ND} further simplifies to
\begin{align}
\label{eq:WZW_ND_emis}
\hspace*{-0.3cm}
\frac{Q_{ \gamma N \to N a}^{\rm WZW, ND}}{10^{32}\;\text{erg/s/cm}^3}  \approx & \; 2.2\;  
g_{10}^4\; T_{40}^8 \,\rho_{15} \left(\frac{C_A 10^9}{f_a/\text{GeV}}\right)^2\;,
\end{align} 
here we define $T_{40}$ as $(T/40~\text{MeV})$ and $g_{10}\equiv(g_\omega/10)$. The temperature dependence in Eq.~\eqref{eq:WZW_ND_emis} can be understood as the phase space scales as $T^3$, the cross-section and emitted energy as $T^4$ and $T$ respectively.

\subsubsection{Degenerate scenario}The derivation of the cooling rate is more involved when neutrons are degenerate inside a medium. We closely follow the derivation of Ref.~\cite{Chakraborty_2023} (see also Appendix~\ref{app:computa_em}) to write the emissivity in terms of the momentum transfer and obtain, 
\begin{align}
\frac{Q_{ \gamma N \to N a}^{\rm WZW, D}}{10^{32}\text{erg/s/cm}^3}  \approx 0.66\;  g_{10}^4 T_{40}^9 \bigg(\frac{C_A\;10^9}{f_a/\text{GeV}} \bigg)^2  \,. 
\end{align}
The final emissivity is given by the minimum of the degenerate and the non-degenerate expressions: $Q^{\rm WZW}=\text{Min}[Q^{\rm WZW}_{\rm D}, Q^{\rm WZW}_{\rm ND}]$. At the saturation density $\rho\sim \rho_0 \approx 0.3 \times 10^{15} \text{g cm}^{-3}$, we obtain that the rate is 
\bea
    Q^{N \gamma \to N a} = \begin{cases}
      Q^{N \gamma \to N a}_{\rm D}   \qquad T \lesssim 40 \text{MeV}
      \\
       Q^{N \gamma \to N a}_{\rm ND}   \qquad T \gtrsim 40 \text{MeV}
    \end{cases}
\eea 
and we conclude that for the SN, it suffices to use the ND limit. On  Fig.~\ref{fig:comparison2}, we show the emissivity of the photo-production and compare it with pion conversion and bremsstrahlung.

At this point, it is worth mentioning that the high densities in the SN cores lead to the modification of the masses and the couplings. In Appendix \ref{app:uncertainties}, we find a very modest in medium uncertainties
\bea 
\frac{Q^{\star}_{N\gamma \to Na}}{Q_{N\gamma \to Na}}\approx  \frac{m_\omega^4}{ g_\omega^4}  \frac{g_{\omega\star}^4  }{m_{\omega\star}^4} \in [0.8, 1.5]\;.
\eea 

From those considerations, it is clear that cooling from the WZW term is subdominant by at least an order of magnitude compared to the prototypical processes as shown in Eq.~\eqref{Eq:bound_WZW}. Although the result strongly depends on the effective coupling $g_\omega$, we note that it is constrained by nucleon phase-shift and photon-proton scattering data~\footnote{We thank Andrea Caputo for pointing those available data to us.}. Therefore, in this work, we consider it to be $g_\omega\sim g_\pi$.

\subsection{Data-driven approach of axion photo-production} In this section, we take a complementary approach and propose a data-driven way of probing the axion parameter space. Recently the PIONS@MAX-lab Collaboration~\cite{Strandberg:2018djk} measured the total photo-production cross-section of pions near the threshold. The collected data of $p \gamma \to p \pi_0$ in the region $E_\gamma=145-180$ MeV~\cite{Briscoe:2020qat} fits well with the predictions from SAID MA19~\cite{A2:2019yud} and MAID2007~\cite{Drechsel:2007if}. 

\begin{figure}[t]
\centering
  \includegraphics[width=\linewidth]{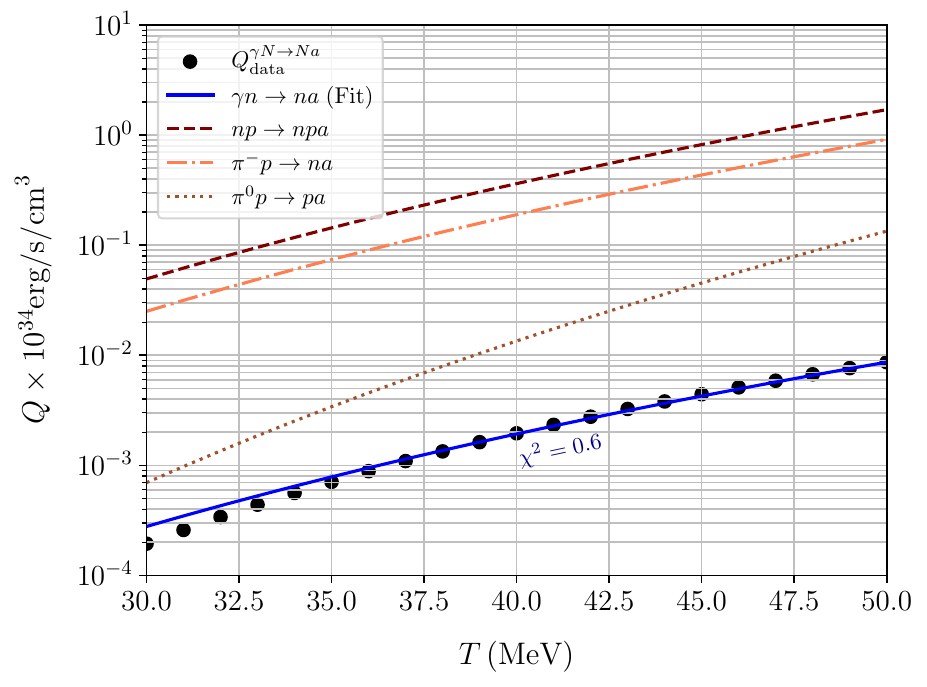}
 \caption{Comparison of the emissivity rates from axion bremsstrahlung and pion conversion from \cite{Choi:2021ign,PhysRevD.64.043002} with WZW photo-production (blue line). The black dot points are the numerical estimates of the emissivity integral, while the blue line is the fit presented in Eq.\eqref{eq:fit_data}. For photo-production, we find that the data-driven approach contributes dominantly.}
 \label{fig:comparison2}
 \end{figure}

Similarly, the A2 collaboration at MAMI reported a new and precise determination of the differential and total cross section for $p \gamma \to p \pi_0$ in the energy range  $E_\gamma=180-500$~\cite{PhysRevC.100.065205}. We fit the pion photo-production cross section using these data points, as shown in Fig.\ref{fig:FIT}. This, in turn, allows us to determine the axion photo-production cross-section, i.e., $\sigma_{\gamma N\to N a}$ in an entirely data-driven way. Integrating from threshold $E_\gamma \in[145, 500]$ MeV, we obtain a characteristic fit for the axion emissivity as
\begin{align}
\label{eq:fit_data}
\frac{Q_{\rm data} }{  10^{33} \text{ cm}^{-3}s^{-1}\text{erg}} \approx   1.54 \bigg(\frac{C_A  10^9}{f_a/\text{GeV}} \bigg)^2  \rho_{15}\;  T_{40}^{6.73} \,, 
\end{align}
 where the exponent of the temperature $6.73$ provides a good fit to the numerical evaluation with $\chi^2$ of 0.6 for Temperatures $T \in [30, 50]$ MeV and is presented in Fig.\ref{fig:comparison2}.  Clearly, the data-driven contribution to axion emissivity dominates over the WZW term. In the regime where data is not present, i.e., $E_a \in [0, m_\pi]$, we compute the emissivity via the WZW term alone. This emissivity is compared with other sources of axions during SN in Fig.\ref{fig:comparison2}.

\subsection{Additional suppressing effects:} 
We now discuss the inclusion of subtleties that might modify the cooling computation, namely, the absorption in the core and the lapse effect.
\subsubsection{Absorption and decay inside the SN core}
The cooling argument in Eq.~\eqref{Eq:bound_WZW} applies only when the axions can free-stream through the SN core and do not get trapped\cite{PhysRevD.42.3297}. Absorption proceeds via $a N \to \gamma N$ and $NNa \to NN$,~\cite{Carenza:2019pxu, Lella:2022uwi, Lella:2023bfb}. For the mean free paths we obtain
\bea 
\frac{L^{apn \to pn}_a}{10\;\text{Km}} \approx  \frac{10^{2}}{X_p C_{ap}^2}  \bigg(\frac{f_a}{10^{9} \text{GeV} }\bigg)^{2} \frac{E_a}{1\text{ GeV}} \, ,
\eea

and
\begin{equation}
    \frac{L^{a N \to N \gamma}_a}{10\;\text{Km}} \approx \frac{\rho_{15}^{-1}}{C_A^2}\;\frac{250}{ g_{10}^4}\left(\frac{f_a}{10^9~\text{GeV}}\right)^2 \left(\frac{1~\text{GeV}}{E_a}\right)^4\;.
\end{equation}
This translates into a lower and an upper bound on the energy of the escaping axion. Therefore, Eq.~\eqref{eq:ND} was an estimate based only on free-streaming axions. We impose these cut-offs on the axion energy while deriving the final limits on the axion parameter space, shown on Fig.~\ref{fig:comparison}.

Moreover, if the axion is massive, it can decay to two photons due to the unavoidable coupling $\mathcal{L} \subset G_{a \gamma \gamma} a \vec{E} \cdot \vec{B}$. 
The rate for this decay is
\cite{Caputo:2022mah} 
\begin{align} 
L_{a \to \gamma \gamma} &= \frac{64\pi}{G_{a \gamma \gamma}^2} \frac{\sqrt{E_a^2- m_a^2}}{E_a^4} 
\nonumber \\
&\approx \frac{4 \times 10^4 \text{ km}}{(G_{a \gamma \gamma}/10^{-9}\text{GeV}^{-1})^2} \frac{E_a/100\text{ MeV}}{(m_a/100 \text{ MeV})^4} \, . 
\end{align} 
Since $G_{a \gamma \gamma} \sim 1/(4\pi f_a) $ (which is generated at loop level in our minimal model),  we conclude that for $f_a \gtrsim 10^7$ GeV and $m_a < 300$ MeV, the axion is always stable on the scale of the core. We will not consider this possibility anymore.
\subsubsection{Lapse effects}
The proto-neutron star is a very dense environment and gravitational effect will impact the emissivity via mostly two mechanisms:
\begin{enumerate}
\item The overall energy of the emitted axions is redshifted, 
\item Heavy axions will be trapped\cite{Lucente:2020whw, Lella:2022uwi, Lella:2023bfb}.
\end{enumerate}
To account for this effect, we introduce a lapse factor $\alpha(M, r) \simeq \sqrt{1- 2M/r}$\footnote{We caution the reader that this expression neglects contributions from the pressure and energy of the stellar medium.}, and the emissivity gets modified to\cite{Caputo:2022mah} 
\bea 
 Q^{\text{ lapse effects }}_{N\gamma \to N a} \approx   \int_{\frac{m_a}{\alpha}} \frac{ d E_\gamma }{\pi^2 } E_\gamma^3 f_\gamma (p_\gamma) n_B \sigma \times \alpha^2 \;, 
\eea 
where $\alpha \equiv \alpha(M_{\rm SN}, R_{\rm SN})$. For $R_{\rm SN} = 10$ km and $M_{\rm SN} =  M_{\rm sun}$, $\alpha \approx 0.84$. 

\subsection{Implications of the photo-production} 
\begin{figure}[t]
\centering
  \includegraphics[width=\linewidth]{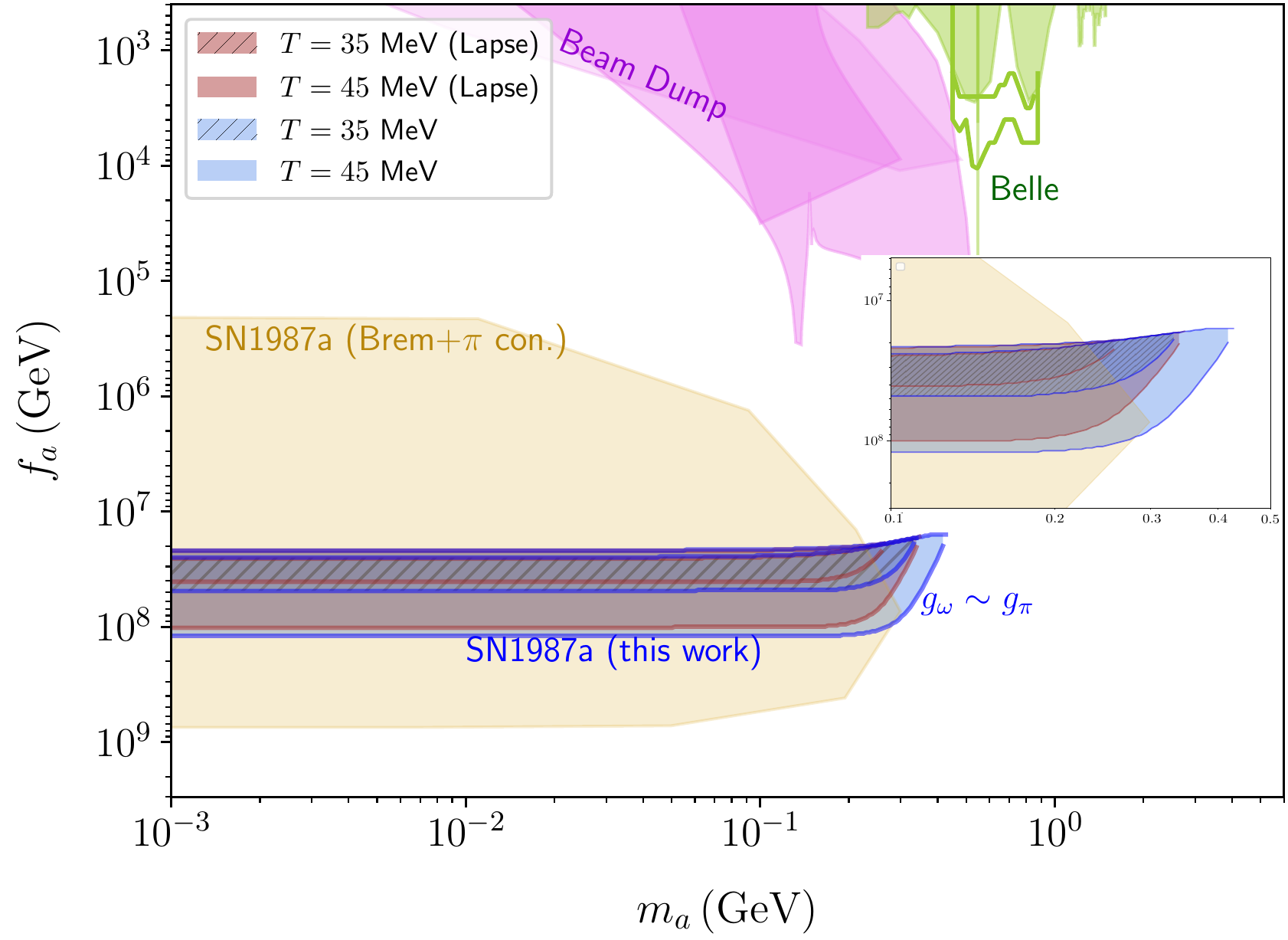}

 \caption{Final exclusion plot due to the photo-production. The red and blue region corresponds to limits from the emissivity from photo-production processes. The blue region is devoid of any lapse correction while the red region takes it into account with ($\alpha = 0.84$).  For all curves, we considered a KSVZ axion model with $C_A \approx 0.2$. Exclusion limits from beam dump experiments~\cite{Bjorken:1988as,Blumlein:1990ay,CHARM:1985anb}, rare decays of  $K$~\cite{Gori:2020xvq,Bauer:2021wjo} and $B$-mesons~\cite{Aloni:2018vki,Chakraborty:2021wda,Bertholet:2021hjl} are shown in magenta and green respectively. We recast the previous bounds~\cite{Lella:2023bfb} on axion parameter space from bremsstrahlung and pion conversion in yellow. Limits on effective axion-photon coupling $g_{a\gamma\gamma}$~\cite{Agrawal:2021dbo,Graham:2015ouw}, from HB stars~\cite{Ayala:2014pea,Straniero:2015nvc}, CAST~\cite{CAST:2017uph}, Sumico~\cite{Graham:2015ouw} and light shining through wall~\cite{Redondo:2010dp} are important but not relevant for our region of interest.}
 \label{fig:comparison}
 \end{figure}

We summarize our results in fig.~\ref{fig:comparison} where we show that photo-production alone can exclude regions, if $m_a \lesssim 100$, with  $f_a \lesssim  (0.4, 1)\times 10^{8}$ GeV for $T\sim (35, 45)$ MeV respectively. This is however subleading to the previous constraints, deduced from simulations in\cite{Lella:2023bfb}, which reaches $f_a \lesssim 0.77\times 10^{9}$ GeV for $m_a \lesssim 50$ MeV in most of the parameter space. The trapping region beyond $f_a \lesssim 10^7$ can in principle be probed through detailed simulations including both the WZW and the bremsstrahlung type of interaction, however, it is beyond the scope of our present analysis. We also attract attention to the corner of large $m_a$ axions, which might exceed the previous bounds. Even if they have been introduced in a phenomenological way in this study, a more complete inclusion of lapse effects and decay of the axion in flight might influence this conclusion and would need further numerical investigation. 

It is also important to mention that the landscape of QCD axion models is quite broad. Without going into the details of such different scenarios, we consider a minimal situation given by Eq.~\eqref{eq:axion_Lag}, similar to the KSVZ~\cite{Kim:1979if,Shifman:1979if} framework. However, we find that similar conclusions hold in other types of axion models such as DFSZ~\cite{Zhitnitsky:1980tq,Dine:1981rt}.  

\subsection{Emitted Spectrum} The emissivity is related to the number density of particles emitted via
\bea 
\label{Eq:N_p_vs_Q}
Q_{\rm a} = \int dE_a E_a \frac{d^2 n_a}{dE_a dt} \,,
\eea 
where the spectrum is defined as the number of axions emitted per interval of energy and time. Comparing with Eq.~\eqref{eq:emissivity_exp}, it is straightforward to obtain the spectrum in the ND limit and for $m_a \ll T$. Furthermore, the total emitted flux is well approximated by integrating the spectrum over the radius of the star in the optically thin limit. Taking $R_{\rm PNS} \approx 10$ Km, we obtain, in the range $E_a \in [0, 145]$
\begin{align}
\frac{d^2 N^{\rm WZW}_a}{dE_a dt} &\approx C^{\rm WZW}_f  \rho_{15} \bigg(\frac{C_A 10^9}{f_a/\text{GeV}}\bigg)^2g^4_{10} \bigg(\frac{E_a}{\text{MeV}}\bigg)^6 e^{-E_a/T}\;,
\nonumber \\ 
C^{\rm WZW}_f &= 1.8\times 10^{40}\; \text{MeV}^{-1}\text{s}^{-1} \,,
\end{align}
and in the range $E_a \in [ 145, \infty]$, 
\begin{align}
\frac{d^2 N^{\rm data}_a}{dE_a dt} &\approx C^{\rm data}_f \rho_{15} \bigg(\frac{C_A 10^9}{f_a/\text{GeV}}\bigg)^2 \frac{\sigma^{\rm data}_{\gamma n \to n \pi_0} }{\mu b}
\bigg(\frac{E_a }{\text{MeV}}\bigg)^2  e^{-E_a/T} 
\nonumber\\
C^{\rm data}_f &\approx 8.4 \times 10^{48} \text{MeV}^{-1}\text{s}^{-1}   \,.
\end{align}

\begin{figure}[!h]
\centering
  \includegraphics[width=\linewidth]{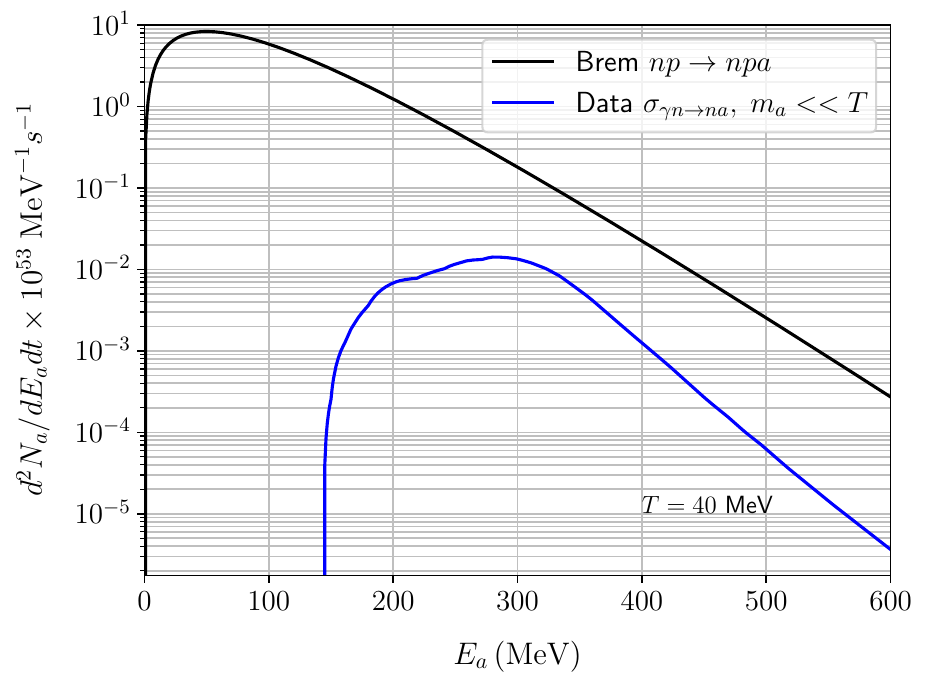}
 \caption{
Spectrum of axions due to WZW photo-production (blue) and bremsstrahlung~\cite{Raffelt_2001} (black). We observe that the number of axions from photo-production is always at least one order of magnitude below the contribution from bremsstrahlung.}
 \label{fig:Spec}
 \end{figure}

As shown in Fig.~\ref{fig:Spec} by the blue curve, the total flux from the WZW photo-production is distinctively different from axion bremsstrahlung, with a comparatively harder spectrum. 

\subsection{Axion detection prospects}
Finally, we discuss the prospect of detecting such axions in Water Cerenkov experiments~\cite{PhysRevLett.65.960} such as Hyper-Kamiokande with fiducial mass 374 ktons \cite{Abe:2011ts}. Usually, axions emitted during future galactic SN can interact with the detector via the channels $ap \to p \gamma$ and $a p \to n\pi^+$~\cite{Lucente:2022vuo}. Around the peak of the WZW axion spectrum, i.e., $E_a \sim 200-300$ MeV, the cross-section for $a p \to n\pi^+$ is dominated by the $\Delta(1.23~\text{GeV})$ resonance and roughly translates to $\sigma_{a p \to N \pi^+} \approx 10^{-25}C_{A}^2 (f_{\pi}/f_a)^2$ cm$^{-2}$~\cite{2007suph.book.....H}. Assuming $N_t$ target particles, we evaluated the number of pions produced during such interaction can be obtained:
\bea
\frac{dN_\pi}{dt} \approx  
0.2\;\rho_{15}\; C_A^4\;  \bigg(\frac{10^9}{f_a/\text{GeV}}\bigg)^4 \bigg(\frac{\text{Kpc}}{d}\bigg)^2
\nonumber \\
\times  \bigg(\frac{M_{\rm detector}}{\text{kton}}\bigg) \bigg(\frac{\text{g}/\text{mol}}{m_{H_2 O}}\bigg)  \, ,
\eea 
where $m_{H_2 O} = 18 \text{g}/\text{mol}$ the molecular mass of water. 
A SN at a distance of 0.2 Kpc with axions having $f_a = 10^9$ GeV, would only roughly produce $\sim 1$ pion in the Hyper-Kamiokande, on a timescale of 10 seconds. This production is thus likely negligible.

\section{Summary \& outlook}
\label{sec:conclusion}
In summary, we study in detail the emissivity during supernovae originating from axion photo-production, i.e., $\gamma N\to N a$. We use the experimentally obtained scattering rates available in the range $E_\gamma\sim E_a \subset [145, 500]$ MeV together with the contribution from the WZW term $\epsilon^{\mu\nu\alpha\beta} F_{\mu\nu}\;\partial_\alpha a\;\omega_\beta$. We find that photo-production is subdominant with respect to the usual axion bremsstrahlung process $NN\to NNa$, as depicted in Fig.~\ref{fig:comparison} in most of the parameter space. It could however bring relevant corrections in the range of massive axions $m_a \gtrsim 100$ MeV, but we emphasize that such a claim would require further numerical computations~\cite{guptaSim}. Moreover, the spectrum of such axions is distinctively harder compared to the prototypical SN cooling processes, with a peak in the spectrum around $E_a \sim (5-6)T$.  

Alternatively, photo-production of axions can also be probed at GlueX\cite{Aloni:2019ruo}, where the photon beam of $\sim 10$ GeV hits a fixed target and produces a plethora of mesons. As far as degenerate neutron stars are concerned, we find that the cooling rate scales as $T^9$ and is subdominant compared to $NN \to NNa$.

\section*{Acknowledgements} SC thanks the Science and Engineering Research Board, Govt. of India (SRG/2023/001162) for financial support. MV is supported by the ``Excellence of Science - EOS" - be.h project n.30820817, and by the Strategic Research Program High-Energy Physics of the Vrije Universiteit Brussel. AG is supported by the “Generalitat Valenciana" and CSIC through the GenT Excellence Program (CIDEIG/2022/22). MV thanks IFIC (Valencia) and CSIC for their hospitality during the completion of this project and Alberto Mariotti for comments on the draft. Authors would also like to thank Andrea Caputo, Stefan Stelzl, Konstantin Springmann and Michael Stadlbauer for very helpful discussions on the draft.

\appendix

\section{Wess-Zumino-Witten Interactions}
\label{App:WZW}
The dynamics and interactions of Goldstone bosons associated with the spontaneous breaking of chiral symmetry is aptly described by the Chiral Lagrangian ($\chi$PT). At the leading order, the Lagrangian consists of terms 
\begin{equation}
    \mathcal L = \frac{f_\pi^2}{4} \text{Tr}\left(D_\mu U^{\dagger} D^\mu U+ U^{\dagger}\chi+\chi^\dagger U\right)\;,
    \label{eq:Chiral_Lagrangian}
\end{equation}
where $D_\mu U = \partial_\mu U -i r_\mu U + i U \ell_\mu$. $r_\mu$ and $\ell_\mu$ describe left and right-handed currents which can be used to construct the vector and axial vector currents. Moreover, $\chi=2B_0\mathcal M$, where $\chi$ is the isospin symmetry breaking spurion with $\mathcal M$ denoting the quark mass matrix. Constants $B_0$ and $f_\pi$ are fixed by experiments. However, Eq.~\eqref{eq:Chiral_Lagrangian} possesses more symmetries such as spurious parity and charge conjugation compared to its underlying UV theory, i.e., QCD. As a result, $\chi$PT in Eq.~\ref{eq:Chiral_Lagrangian} fails to describe processes such as $K^+K^-\to \pi^+\pi^-\pi^0$ and $\pi^0\to\gamma\gamma$, which are otherwise allowed in QCD. Such interactions can be restored in the Chiral Lagrangian by the Wess-Zumino-Witten term. In the geometric representation of the WZW form, one considers a 5-dimensional action, whose boundary is identified with 4-dimensional space in the form
\begin{eqnarray}
   && S_{WZW}(U) = \kappa \int_D d^5 y\; \omega\;,\; \text{where}\;\nonumber\\
   && \omega=-\frac{i}{240\pi^2} \epsilon^{\mu\nu\rho\sigma\tau}\;\text{Tr}\left[\;\mathcal{U}_\mu\; \mathcal{U}_\nu\; \mathcal{U}_{\rho}\; \mathcal{U}_\sigma\; \mathcal{U}_\tau\right]\;.
    \label{eq:Saction}
\end{eqnarray}
Here, we define $\mathcal{U}_\mu=U^{\dagger}\partial_\mu U$ and $\mathcal{U}\approx \text{Exp}\left(2i\pi^a T^a/f_\pi\right)$ with $f_\pi=93$ MeV. Under chiral symmetry, $U$ transforms as $U\to R U L^\dagger$ with $R$ and $L$ describing the $SU(2)_R$ and $SU(2)_L$ transformation matrices. Note that Eq.~\eqref{eq:Saction} correctly describes the process $K^+K^-\to\pi^+\pi^-\pi^0$. To generate $\pi^0\to\gamma\gamma$ or similar processes, one has to gauge $S_{WZW}$. However, any gauging has to be done in the 4-dimension. In principle, any arbitrary subgroup of the WZW action can be gauged and the resultant terms are 
\begin{eqnarray}
        S_{WZW} \left(U,A_{L},A_{R}\right) &=& S_{WZW}\left(U\right)\nonumber \\
        &+&\frac{N_c}{48\pi^2}\int d^4 x\; \epsilon^{\mu\nu\rho\sigma}\; Z_{\mu\nu\rho\sigma}\;,
\end{eqnarray}
where the full expression of $Z_{\mu\nu\rho\sigma}$ has been nicely tabulated in Ref.~\cite{Witten:1983tw}. However, for our work, we only track the pion fields and therefore the relevant terms are
\begin{widetext}
    \begin{align}
    Z_{\mu\nu\rho\sigma} (U, A_L, A_R)\supset i\; & \text{Tr} \left[\left\{\left(\partial_\mu A_{\nu_L}\right)A_{\rho_L}+A_{\mu_L}(\partial_\nu A_{\rho_L})\right\}\partial_\sigma U\; U^{-1}+\left\{\left(\partial_\mu A_{\nu_R}\right)A_{\rho_R}+A_{\mu_R}(\partial_\nu A_{\rho_R})\right\}U^{-1}\partial_\sigma U\right]\nonumber \\
    +& i\; \text{Tr} \left[\left\{(\partial_\mu A_{\nu_R})\;U^{-1}A_{\rho_L}\partial_\sigma U+A_{\mu_L} U^{-1} (\partial_\nu A_{\rho_R})\;\partial_\sigma U\right\}\right]\;.
    \label{eq:Z}
\end{align}
\end{widetext}

Expanding in pion fields, $\partial_\mu U$ can be written as
\begin{equation}
\partial_\mu U = i\frac{\partial_\mu \pi_0}{f_\pi}   \begin{pmatrix}
    1 & 0
    \\ 0 & -1
\end{pmatrix} + ... \, .
\end{equation}
This choice results in a canonically normalized kinetic term from $\pi_0$. Moreover, in the SM, the gauge fields are expressed as
\begin{align}
  A_L = g_2 W^a \frac{\tau^a}{2} + g_1 W^0\; 
  \begin{pmatrix} 
  \frac{1}{6} & 0 \\
  0 & \frac{1}{6} 
  \end{pmatrix}\;, \\
  A_R = g_1 W^0\; 
  \begin{pmatrix}
  \frac{2}{3} & 0 \\
  0 & -\frac{1}{3}
  \end{pmatrix}\;.
  \label{eq:gauge_fields}
\end{align}
The physical gauge fields $A$ and $Z$ are introduced after electroweak symmetry breaking and they are related with $W^0$ and $W^3$ fields via the weak rotation angle $\theta_w$ as
\begin{equation}
W^0 = - s_W Z + c_W A, \qquad W^3 =  c_W Z + s_W A \,. 
\end{equation}
Using $g_1  c_W = g_2s_W= e$, Eq.~\eqref{eq:gauge_fields} becomes
\begin{eqnarray}
A_L &&= g_2s_W\frac{ A}{2}  \begin{pmatrix}
    1 & 0
    \\ 0 & -1
\end{pmatrix}
+ g_1 c_W A \begin{pmatrix}
    \frac{1}{6} & 0
    \\ 0 & \frac{1}{6}
\end{pmatrix}\;,\\
&&= eA\begin{pmatrix}
    \frac{2}{3} & 0
    \\ 0 & -\frac{1}{3}
\end{pmatrix} = A_R  \, .
\end{eqnarray}

To introduce the vector mesons fields, we use the background field method presented in Ref.~\cite{Harvey:2007ca} where $\widetilde{A}_{L,R}=A_{L,R}+B_{L,R}$ (the charge as been included inside the definition of the field), including both the fundamental and background gauge fields. As a result, we make the following replacement $A_{L,R}\to\widetilde{A}_{L,R}$ in Eq.~\eqref{eq:Z}. The background field $\omega$ is then identified as 
\begin{align}
   B_V =  \frac{2g_\omega}{3}\; \omega \begin{pmatrix}
    1 & 0
    \\ 0 & 1
\end{pmatrix} \, ,
\qquad 
B_A = 0\, , \\
\Rightarrow  B_L = B_R = \frac{B_V}{2} = \frac{g_\omega}{3}\; \omega \begin{pmatrix}
    1 & 0
    \\ 0 & 1
\end{pmatrix}\,. 
\end{align}
Using the definition of $A_{L,R}$ and $B_{L,R}$ as well as $N_c=3$, fixed by matching with QCD, we find from  Eq.~\eqref{eq:Z}
\begin{eqnarray}
\label{eq:vertex_WZW_pi}
    \mathcal{L}_{\rm WZW} &\supset & \frac{N_c}{24\pi^2}e g_\omega \frac{\partial_\rho \pi_0}{f_\pi} \epsilon^{\mu\nu\rho\sigma} F_{\mu\nu}\; \omega_\sigma \nonumber \\
    &+& \frac{N_c}{96\pi^2}e^2 \; \epsilon^{\mu\nu\rho\sigma}\; \frac{\pi_0}{f_\pi} F_{\mu\nu} F_{\rho\sigma}\;.
\end{eqnarray}
The coefficient $N_c=3$ is fixed by matching with QCD.

\section{Computation of the emissivities for the axion}
\label{app:computa_em}
In this section, we present the computation of the emissivities of the WZW photo-production reaction $\gamma N \to Na$ for the non-degenerate case and then for the degenerate case. 
\subsection{Emissivity in the non-degenerate case}
We study the $2 \to 2$ interaction $\gamma N \to N a$ in the non-degenerate limit. In the centre of mass frame of the collision, $E_{CM}= E_{N_1}+E_\gamma = E_{N_2}+E_a$. 
The matrix element takes the form 
\begin{align}
\langle |\mathcal{M}|^2\rangle & = 8\frac{\kappa^2  m_N^2 g_\omega^2}{f_a^2m_\omega^4}  \bigg(E_\gamma^2 |\vec{p}_a|^2 \sin^2 \theta  \bigg) + \mathcal{O}(p_F^2/m_N^2)\;,
 \end{align}
with $\kappa \equiv C_{A}g_\omega e N_c/24\pi^2$. We also neglect $m_\gamma$ compared to the energy scale. $\theta$ is the angle between the incoming neutron and the outgoing axion and $p_F$ is the Fermi momentum of neutrons. Also, $E_i$ and $\vec{p_i}$ are the energy and three-momenta of the $i^{\rm th}$ species. In the CM frame $E_{\rm CM} \equiv E_{N_1} + E_\gamma = E_{N_2} + E_a$,  $E_\gamma \approx E_a$ and $E_a^2 = p_a^2 + m_a^2$.
The cross-section for the $N\gamma\to N a$ process is given by 
\begin{align}
\frac{d\sigma_{N\gamma\to Na}}{d\Omega} &=\frac{|\vec{p}_a|}{|\vec{p}_\gamma|}\frac{\langle|\mathcal{M}|^2\rangle}{64\pi^2 E_{CM}^2}\theta\left(E_{cm}-m_N-m_a\right)\;, \nonumber \\
&\approx \frac{\sqrt{E_a^2 - m_a^2}}{E_\gamma}\frac{\langle |\mathcal{M}|^2\rangle}{64 \pi^2 m_N^2}  \theta(E_{CM} - m_N - m_a)  \,. 
\end{align}

The emissivity rate in the non-degenerate limit is expressed as
\begin{eqnarray}
 Q^{\gamma N \to N a}_{\rm ND}  \approx  &&\int \frac{g_\gamma d^3 \vec{p}_\gamma}{(2\pi)^3 }  f_\gamma (p_\gamma) \int \frac{g_N d^3 \vec{p}_{N}}{(2\pi)^3 }  f_N \int 
 \frac{\langle |\mathcal{M}|^2\rangle}{4E_{N_1}E_\gamma}\nonumber \\
 &&(2\pi)^4 \delta^4\left(\sum_i p_i\right)\frac{d^3\vec{p_2} d^3\vec{p_a}}{(2\pi)^6 4E_aE_{N_2}}E_a \,. 
\end{eqnarray} 
which can be further simplified with
\begin{eqnarray}
\label{eq:emissivity_bis}
&& Q^{\gamma N \to N a}_{\rm ND}  \approx   \int \frac{g_\gamma d^3 \vec{p_\gamma}}{(2\pi)^3 }  f_\gamma (p_\gamma) \int \frac{ g_N d^3 \vec{p}_{N}}{(2\pi)^3 }  f_N \sigma (E_\gamma) E_a\;, \nonumber \\
&&\sigma(E_\gamma) \approx \frac{1}{3 \pi } \frac{\kappa^2 g_\omega^2}{f_a^2 m_\omega^4} \sqrt{1 - \frac{m_a^2}{E_\gamma^2}}   E_\gamma^2 (E_\gamma^2 - m_a^2) \,, 
\end{eqnarray}
where $\sigma(E_\gamma)$ is the integrated cross section. The nucleon density is given by 
\begin{equation}
\label{eq:density_neu}
 n_B \equiv \int  \frac{d^3 p_{N}}{(2\pi)^3  }  g_N  f_N ,  \qquad n_B = 4.5 \times  10^{6} \rho_{15} \text{ MeV}^3  \,. \nonumber \\
\end{equation}
Plugging Eq.\eqref{eq:density_neu} in the emissivity Eq.\eqref{eq:emissivity_bis}, we obtain
\begin{eqnarray}
\label{eq_Q_non-dege}
Q^{N \gamma \to N a}_{\rm ND} &&\approx  \frac{g_\gamma  n_B}{3 \pi } \frac{\kappa^2 g_\omega^2}{f_a^2 m_\omega^4} \int^{\infty}_{m_a} \frac{d E_\gamma}{2\pi^2} f_\gamma (p_\gamma)  
E_\gamma^7(1 - m_a^2/E_\gamma^2)^{3/2}\, .\nonumber \\
\end{eqnarray}
The two limiting conditions are: 
\begin{enumerate}[leftmargin=*]
\item $m_a/T \to 0$:
In the limit $m_a/T \to 0$, we can approximate $\sqrt{1 - \frac{m_a^2}{E_\gamma^2}} \to 1$. This results in
\begin{align}
Q^{N \gamma \to N a}_{ \rm ND} \approx   \frac{g_\gamma n_B}{3 \pi^3 } \frac{4 \pi^8}{15} \frac{\kappa^2 g_\omega^2}{f_a^2 m_\omega^4} T^8  \approx 27g_\gamma n_B \frac{\kappa^2g_\omega^2}{f_a^2m_\omega^4} T^8\;.
\end{align}

Using the values and the normalization of the parameters explained in the main text along with $g_\gamma = 2$, we get
\begin{eqnarray}
\label{eq:large_T_app}
\frac{Q_{ \gamma N \to N a}^{\rm WZW, ND}}{10^{32}\;\text{erg/s/cm}^3}  \approx & \; 2.2\;  
g_{10}^4\; T_{40}^8 \,\rho_{15} \left(\frac{C_A 10^9}{f_a/\text{GeV}}\right)^2\, .\nonumber \\
\end{eqnarray}

\item $m_a/T \gtrsim 1$:
In the opposite limit $m_a/T \gtrsim 1$, the mass of the axion cannot be neglected anymore and the expression becomes
\begin{eqnarray}
    \frac{Q^{N \gamma \to N a}_{\rm ND} }{4 \times 10^{28}\text{erg/s/  cm}^3}   &&\approx      \bigg(\frac{C_{A}m_N 10^9}{f_a} \bigg)^2g_{10}^4
\times \nonumber \\
&& \rho_{15} G(m_a, T_{40}) \,, 
\end{eqnarray}
where we defined the $G$ function in terms of the Bessel functions as
\begin{eqnarray}
G(m_a, T) 
&&\equiv  3 m_a^4 T^2 \bigg( 4 m_a T K_3\bigg( \frac{m_a}{T}\bigg) \nonumber \\
&&+ (m_a^2 + 35 T^2) K_4\bigg( \frac{m_a}{T}\bigg) \bigg) \,. 
\end{eqnarray}
\end{enumerate}
\subsection{Emissivity in the degenerate case}
\label{appendix:deg}
We now evaluate the emissivity in the presence of strongly degenerate neutrons. The full expression is provided in a convenient form as 
\begin{eqnarray}
&& Q^{N \gamma \to N a}_{\rm axion} = 2\frac{\kappa^2}{f_a^2}\int \dfrac{d^3\vec{p}_\gamma}{(2\pi)^3}
g_\gamma\dfrac{f_\gamma}{2E_\gamma}
\int \dfrac{d^3\vec{p}_a}{(2\pi)^3 2E_a}  \nonumber \\
&\times&\bigg[ 2E_\gamma E_a (p_\gamma \cdot p_a)-(p_\gamma \cdot p_a)^2  \bigg] \, S(q^\mu) E_a  \,. 
\end{eqnarray}

where we define $Q_0 \equiv E_2 - E_1 = E_\gamma - E_a\;,  \vec{q} = \vec{p}_\gamma - \vec{p}_a$. The response function is defined as 
\begin{widetext}
\begin{align}
S(q^\mu) \equiv & g_{N_1}  \int \dfrac{d^3\vec{p}_{N_1}d^3\vec{p}_{N_2} m_N^2}{(2\pi)^6\, E_{N_1}\,E_{N_2}} f_N(E_{N_1})(1-f_N(E_{N_2})) \times (2\pi)^4 \times \delta(E_{N1}+Q_0-E_{N_2})
\delta^3(\vec p_{N_1}-\vec p_{N_2}+\vec q)\,\nonumber \\
&\Theta(E_{N_1}-m_N) \,\Theta(E_{N_2}-m_N)\;,
\label{eq:SDEF2}
\end{align} 
\end{widetext}
and can be computed while identifying $z = Q_0/T$ and $|\Vec{q}|\equiv q$. We obtain the final form of the response function 
\begin{equation}
S(Q_0, q)=\dfrac{M_N^2\,Q_0}{\pi q}\dfrac{1}{1-e^{-z}}\Theta(\mu - E_-).
\label{eq:Final2S}
\end{equation}
\begin{eqnarray}
E^- &=& \sqrt{\bigg(m_N+\frac{Q_0}{2}\bigg)^2 + \bigg(\frac{\sqrt{q^2-Q_0^2}}{2}-\frac{m_NQ_0}{\sqrt{q^2-Q_0^2}}\bigg)^2} \nonumber \\
&-& \left(m_N+\frac{Q_0}{2}\right) \,.
\end{eqnarray}
which can be simplified further

\begin{equation}
\label{Eq:E_-}
E^- \approx  \sqrt{ m_N^2 + \bigg(\frac{m_NQ_0}{\sqrt{q^2-Q_0^2}}\bigg)^2}- m_N \,. 
\end{equation}

To perform the integral over $p_a$, we make the following change of variables 
\begin{eqnarray}
&& q dq = E_a E_\gamma d\cos \theta, \qquad dQ_0 = dE_a\;, \nonumber \\
&& \xrightarrow{} \quad \int\frac{d^3 p_a}{(2\pi)^3 2E_a}=\int\frac{dQ_0 dq q}{(2\pi)^2 2}\frac{1}{E_\gamma}  \, ,
\end{eqnarray}
and the integral becomes 
\begin{eqnarray}
\label{Eq:DEGE_3}
&& Q^{N \gamma \to N a}_{\rm D} = \frac{g_\gamma\kappa^2}{f_a^2}\int \dfrac{d^3p_\gamma}{(2\pi)^3}
\dfrac{f_\gamma}{E_\gamma}
\int \frac{ dQ_0\,  dq }{(2\pi)^2 2E_\gamma} \dfrac{M_N^2\,Q_0}{\pi }\nonumber \\
&&\times \bigg[ 2E_\gamma E_a (p_\gamma \cdot p_a)-(p_\gamma \cdot p_a)^2  \bigg] \dfrac{E_\gamma - Q_0}{1-e^{-Q_0/T}}\Theta(\mu - E_-) \,. \nonumber \\
\end{eqnarray} 

The expression between the parenthesis in Eq.\eqref{Eq:DEGE_3}  can be rewritten
\begin{eqnarray}
&& \bigg[ 2E_\gamma E_a (p_\gamma \cdot p_a)-(p_\gamma \cdot p_a)^2  \bigg] = \bigg[ (E_\gamma E_a )^2-(\vec{p}_\gamma \cdot \vec{p}_a)^2  \bigg]\nonumber \\
&& \approx  \frac{(q^2 - Q_0^2)}{4} \bigg(-(q^2 - Q_0^2) + 4E_\gamma (E_\gamma - Q_0) \bigg) \,.
\end{eqnarray}


To perform the integral, we need to define the boundaries of integration. Kinematics impose the condition that $q$ variable varies from $q \in [Q_0, 2 E_\gamma - Q_0]$. The Theta function in Eq.\eqref{Eq:DEGE_3} combined with the expression for $E_-$ in Eq.\eqref{Eq:E_-} further impose the lower bound on the $q$ region of integration as $b |Q_0|$. In the end, we get  
\begin{eqnarray}
q \in \bigg [b |Q_0|, 2E_\gamma -Q_0 \bigg]\;, &&\qquad  a \equiv (\mu_N/M_N+1)^2-1 \,, \nonumber \\
&& b \equiv \sqrt{\frac{a+1}{a}}    \,. 
\end{eqnarray}
Taking the typical value $\mu_N =300 $ MeV, we find $b\approx 1.56$. We will use those two values as a range of uncertainty for the degenerate density of the proto-NS. This induces that the $Q_0$ integral should be performed between 
\begin{equation}
Q_0 \in \bigg[-\frac{2}{b-1} E_\gamma, \frac{2}{b+1} E_\gamma \bigg] \,. 
\end{equation}

We can further separate the parameter space into the region of 
\begin{equation} 
Q_0 \in \bigg[-\frac{2}{b-1} E_\gamma, 0  \bigg] \,,
 \qquad Q_0 \in \bigg[0, \frac{2}{b+1} E_\gamma  \bigg] \,.
\end{equation}
The integral over $q$ can be trivially performed and by finishing the integration numerically, we obtain 
\begin{align}
\frac{Q_{ \gamma N \to N a}^{\rm WZW, D}}{10^{32}\text{erg/s/cm}^3}  \approx 0.66\;  g_{10}^4 T_{40}^9 \bigg(\frac{C_A\;10^9}{f_a/\text{GeV}} \bigg)^2 \,. 
\end{align} 

\section{Sources of uncertainty of WZW emissivity}
\label{app:uncertainties}
In this subsection, we briefly discuss the main sources of uncertainty associated with the WZW photo-emission of axions. As we will see and as we foresaw in the main text, this essentially comes from the uncertainty on the values of nuclear physics quantities in an extremely dense medium like a SN and the uncertainty on the value of $g_\omega$. We conclude that the uncertainty from the coupling largely dominates the final uncertainty on the emissivity.

\subsection{In medium uncertainties}

 The Brown-Rho scaling dictates that the variation of the mass of the vector mesons follows the neutron mass\cite{PhysRevLett.66.2720}:
 \begin{equation}
 \label{Eq:Brown_rho}
  \frac{f_\pi^\star}{f_\pi}\approx\frac{m^\star_\omega}{m_\omega} \approx \frac{m^\star_N}{m_N}\Rightarrow  \frac{g_\omega^\star}{g_\omega} \approx \frac{g_\pi^\star}{g_\pi}  \approx \frac{g_A^{\star}}{g_A} \,.
\end{equation}
where the $\star$ indicates that the quantity is evaluated in the dense environment of the NS~\cite{MAYLE1989515, Payez:2014xsa, Raffelt_2001}. The $m_N$ dependence is obtained from~\cite{1987NuPhA.473R.760.,Raffelt_2001}. For the $g_A$ dependence, we use two different scaling laws: thick line from\cite{Rho:1985ay,MAYLE1989515} and dashed line from\cite{Voskresensky:2001fd,Fischer:2016boc}, $g_A^\star/g_A \approx 1/(1+(1/3)(m^\star_N/m_N)(\rho/\rho_0)^{1/3})$. In the case of the axion bremsstrahlung, such effects led to only moderate uncertainties in the coupling\cite{MAYLE1988188, MAYLE1989515}. For  $\rho/\rho_0\in[0.5, 1]$, the uncertainties in the WZW photo production amounts to
\begin{equation} 
\frac{Q^{\star}_{N\gamma \to Na}}{Q_{N\gamma \to Na}}\approx  \frac{m_\omega^4}{ g_\omega^4}  \frac{g_{\omega\star}^4  }{m_{\omega\star}^4} \in [0.8, 1.5]\;.
\end{equation} 
We conclude that the uncertainty on the emissivity from the in-medium effect is very mild. 

We however emphasize that the in-medium corrections to the data-driven approach will not scale in this simple way and can receive large corrections. 

\subsection{The value of $g_\omega$}

As we discussed in the main text, one of the main sources of uncertainty of our work comes from the value of the coupling $g_\omega$ itself, since the emissivity scales like $g_\omega^4$ and thus the bound on $f_a$ scales like $g_\omega^2$. To alleviate this uncertainty on the coupling, we adopted a data-driven approach in all the regions where data for the pion-photo-production was available. We then fitted the cross-section and applied the mixing angle to obtain a cross-section for the axion photo-production. In Fig. \ref{fig:FIT} we present the data points used and our fit.

\bibliographystyle{apsrev4-1}
\bibliography{bibliography}




\end{document}